# A Tribute to Ingram Olkin

**Edward I. George**

It is with pleasure and pride that I introduce this special section in honor of Ingram Olkin. This tribute is especially fitting because, among the many profound and far-reaching contributions that he has made to our profession, Ingram Olkin was the key force behind the genesis of *Statistical Science*. As put so eloquently by Morrie DeGroot [1], the founding Executive Editor of *Statistical Science*,

> This journal exists largely because of the efforts of one person, Professor Ingram Olkin of Stanford University. It was he who first realized the importance of a new journal of this type both to the field of statistics and probability and to the IMS, and it was his idea that the IMS should establish such a journal. He has worked tirelessly and creatively during the past few years to bring it into existence. Everyone who shares our belief in the usefulness of *Statistical Science*, or enjoys some of the articles that we publish, owes Professor Olkin a debt of gratitude for his vision and his persuasive talents.

I know I speak for all the past Executive Editors of *Statistical Science* when I say we are deeply grateful to Ingram Olkin for his guidance and his many contributions to the journal over the years.

The four papers in this special section were first presented as part of a session in honor of Ingram Olkin's 80th birthday, organized by Allan Sampson and Leon Gleser for the 2004 Joint Statistical Meetings in Toronto, Canada. Authored by Ingram Olkin's own students, these papers convey the far-reaching community of scholars he has produced, as well as a sense of just some of the now flourishing areas that his many seminal contributions have initiated. I am especially grateful to Leon Gleser for his role in getting this special section under way for *Statistical Science*, and to Allan Sampson who finally managed to get Ingram Olkin in one place long enough to obtain the long-awaited conversation which concludes the section. I am delighted to say that it was well worth the wait!

*Edward I. George is Professor, Department of Statistics, The Wharton School, University of Pennsylvania, Philadelphia, Pennsylvania 19104, USA e-mail: edgeorge@wharton.upenn.edu.*

1